\documentclass[%
 aip,
 amsmath,amssymb,
 reprint,%
]{revtex4-1}

\UseRawInputEncoding
\usepackage{graphicx}% Include figure files
\usepackage{dcolumn}% Align table columns on decimal point
\usepackage{bm}% bold math
\usepackage{tabularx}
\usepackage{bm}%
\usepackage{color}      % use if color is used in text
\usepackage[colorlinks=true,linkcolor=blue,citecolor=blue,urlcolor=blue]{hyperref}%
\usepackage{siunitx}
\usepackage{multirow}

\expandafter\ifx\csname package@font\endcsname\relax\else
 \expandafter\expandafter
 \expandafter\usepackage
 \expandafter\expandafter
 \expandafter{\csname package@font\endcsname}%
\fi
\renewcommand{\thesection}{\arabic{section}}

\renewcommand{\baselinestretch}{1}

\begin{document}

\author{B. C. Johnson}
\email{brett.johnson2@rmit.edu.au}
\affiliation{%
Centre of Excellence for Quantum Computation and Communication Technology, School of Engineering, RMIT University, VIC, 3001, Australia
}%

\author{M. Stuiber}
\affiliation{%
School of Physics, University of Melbourne, VIC 3010, Australia
}%
\affiliation{%
Melbourne Centre for Nanofabrication, VIC 3168, Australia
}%
% 0000-0001-7831-4825

\author{D. L. Creedon}
\affiliation{%
School of Physics, University of Melbourne, VIC 3010, Australia
}%

\author{A. Berhane}
\affiliation{%
School of Physics, University of New South Wales, Sydney, NSW, Australia
}%

\author{L. H. Willems van Beveren}
\affiliation{%
School of Physics, University of Melbourne, VIC 3010, Australia
}%

\author{S. Rubanov}
\affiliation{%
Ian Holmes Imaging Centre, Bio21 Institute, University of Melbourne, VIC, 3010, Australia
}%

\author{J. H. Cole}
\affiliation{%
School of Science, RMIT University, VIC, 3001, Australia
}%

\author{V. Mourik}
\affiliation{%
School of Physics, University of New South Wales, Sydney, NSW, Australia
}%
% \affiliation{%
% Forschungszentrum Jülich, 52425 Jülich, Germany
% }

\author{A. R. Hamilton}
\affiliation{%
School of Physics, University of New South Wales, Sydney, NSW, Australia
}%

\author{T. L. Duty}
\affiliation{%
School of Physics, University of New South Wales, Sydney, NSW, Australia
}%

\author{J. C. McCallum}
% \email{jeffreym@unimelb.edu.au}
\affiliation{%
School of Physics, University of Melbourne, VIC 3010, Australia	
}%

% \noindent B. C. Johnson, brett.johnson2@rmit.edu.au, \\
% ORCID = \href{https://orcid.org/0000-0002-2174-4178}{0000-0002-2174-4178}\\ \\

% M. Stuiber, michael.stuiber@nanomelbourne.com \\
% ORCID = \href{https://orcid.org/0000-0001-7831-4825}{0000-0001-7831-4825} \\ \\

% D. L. Creedon, daniel.creedon@unimelb.edu.au\\
% ORCID = \href{https://orcid.org/0000-0003-2912-3381}{0000-0003-2912-3381}\\ \\

% L. H. Willems van Beveren, laurensw@unimelb.edu.au\\
% ORCID = \href{https://orcid.org/0000-0002-4404-0042}{0000-0002-4404-0042}\\ \\

% A. Berhane, a.berhane@unsw.edu.au\\
% ORCID = \href{https://orcid.org/}{}\\ \\

% S. Rubanov, sergey@unimelb.edu.au\\
% ORCID = \href{https://orcid.org/}{}\\ \\

% J. H. Cole, Jared Cole jared.cole@rmit.edu.au\\
% ORCID = \href{https://orcid.org/0000-0002-8943-6518}{0000-0002-8943-6518}\\ \\

% V. Mourik, vincentmourik@gmail.com \\
% ORCID = \href{https://orcid.org/0000-0003-1522-7403}{0000-0003-1522-7403}\\ \\

% A. R. Hamilton, alex.hamilton@unsw.edu.au\\
% ORCID = \href{https://orcid.org/0000-0001-7484-3738}{0000-0001-7484-3738}\\ \\

% T. L. Duty, t.duty@unsw.edu.au,  \\
% ORCID = \href{https://orcid.org/}{} \\ \\

% J. C. McCallum, jeffreym@unimelb.edu.au\\
% ORCID = \href{https://orcid.org/0000-0002-6692-7728}{0000-0002-6692-7728} \\ \\

% \title{Magnetoresistance in superconducting aluminium-silicon alloy nano-rings}
% \title{Fluxoid quantisation in superconducting aluminium-silicon alloy nano-rings}
\title{Phase transformation-induced superconducting aluminium-silicon alloy rings}

\begin{abstract}

% Magnetic flux quantization in superconducting rings is a property of both fundamental and technological importance. 

The development of a materials platform that exhibits both superconducting and semiconducting properties is an important endeavour for a range of emerging quantum technologies. We investigate the formation of superconductivity in nanowires fabricated with silicon-on-insulator (SOI). Aluminium from deposited contact electrodes is found to interdiffuses with the Si nanowire structures to form an Al-Si alloy along the entire length of the predefined nanowire device over micron length scales at temperatures well below that of the Al-Si eutectic. The resultant transformed nanowire structures are layered in geometry with a continuous Al-Si alloy wire sitting on the buried oxide of the SOI and a residual Si cap sitting on top of the wire. The phase transformed material is conformal with any predefined device patterns and the resultant structures are exceptionally smooth-walled compared to similar nanowire devices formed by silicidation processes. The superconducting properties of a mesoscopic AlSi ring formed on a SOI platform are investigated. Low temperature magnetoresistance oscillations, quantized in units of the fluxoid, $h/2e$, are observed.

\end{abstract}

\keywords{Superconductivity, nanowires, fluxoid quantization, alloy formation}

\maketitle
 
% \section{Introduction}

%Devices that integrate solid state spins with superconducting
 %connecting superconducting circuits with solid-state spins 

%R. Ams€uss, C. Koller, T. N€obauer, S. Putz, S. Rotter, K. Sandner, S. Schneider, M. Schramb€ock, G. Steinhauser, H. Ritsch et al., Phys. Rev. Lett. 107, 060502 (2011).

The integration of superconductors within silicon devices is expected to greatly broaden opportunities in emerging quantum technologies.\cite{Shim2014,Blase2009}  Such hybrid devices are promising candidates to explore topological superconductivity and look for the possible existence of Majorana Fermions.\cite{Ridderbos:2020wo, Leijnse2012} Further, unique architectures to couple spin qubits to superconducting microwave resonators\cite{Kubo2010,Schrambock2011} for qubit coupling over long distances may also be possible.\cite{Tosi2017}

Approaches to obtain superconductivity in silicon-related materials and devices have generally relied on high concentration doping with boron, formation of alternative superconducting phases such as certain silicides, or proximity induced superconductivity.\cite{Bustarret2015,Blase:2009tg,Huffelen1993} Two general approaches to render Si superconducting are phase formation and a range of non-equilibrium techniques\cite{Shim2014}. The non-equilibrium doping approach aims to incorporate an extremely high active boron concentration with gas-immersion laser doping \cite{Marcenat2010} or ion implantation followed by annealing\cite{Skrotzki2010,Skrotzki2011,Herrmannsdorfer2009}. In contrast, the phase formation approach is commonly although not exclusively based on metal deposition followed by a thermal anneal to induce silicide or alloy formation. A common advantage of this latter approach is the ease with which it can be implemented and that it is a common process in the formation of Ohmic contacts in microelectronics. Several superconducting silicon based compounds have been investigated using this process such as MoSi\cite{Lehtinen2017}, PtSi\cite{Oto1994} and VSi\cite{Zhang2021}. More recently Al, which is a common choice for superconducting devices with its large coherence length, has been integrated with SiGe\cite{Ridderbos2018} and Ge \cite{Ridderbos:2020wo, Luong2020}. Some of the main advantages of the phase transformation method are the ability to scale the process, to have full control over the cross-sectional area of the device as well as its aspect ratio \cite{Lehtinen2017} and to form intimate contact between the superconducting and the semiconducting components of the device. 
	
% as successfully demonstrated with Nb?? - Si, TaSi, PtSi, VSi, WSi

% Here, we describe a Si nanowire device fabrication process formed on a silicon-on-insulator (SOI) platform. During a thermal in-diffusion anneal of deposited Al contact pads, Al is found to migrate over micron length scales at temperatures below the eutectic temperature. AlSi is formed throughout the nanowire device conforming to the geometry of the initial Si device. Finally, excess Si is pushed to the top of the nano-wire, forming a Si ribbon. Low temperature magnetoresistance measurements on such a device are presented. 

\begin{figure*}
\begin{center}
\rotatebox{0}{\includegraphics[width=16cm]{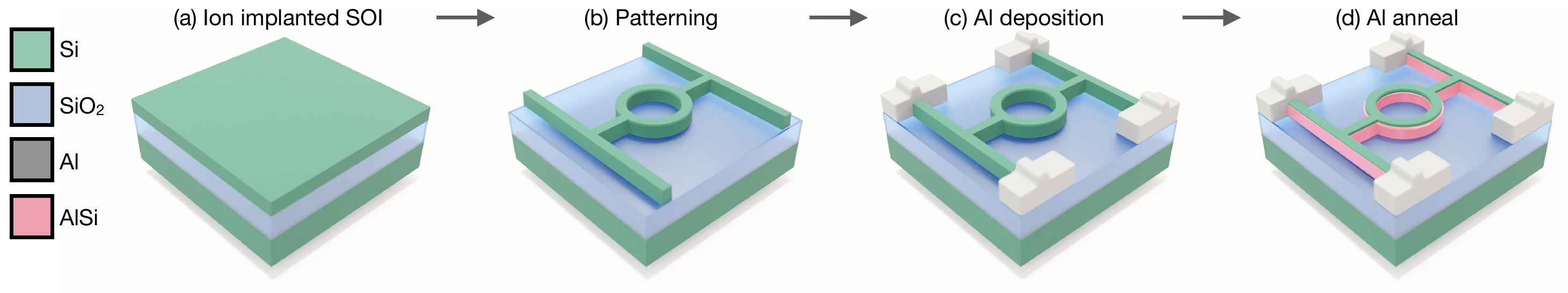}}
\end{center}
\caption[]{Schematic of the fabrication process used to form AlSi nanowires. (a) SOI with a highly doped 50~nm top layer is used.  (b) Si nanowire patterning is achieved with EBL and RIE. (c) Al is deposited over the Si contact pad areas. (d) The final 450$^\circ$C anneal induces migration of the Al.} 
\label{Fig1}
\end{figure*}

In this work, we describe a Si nanowire device fabrication process formed on a silicon-on-insulator (SOI) platform. Electrical contact pads were formed by aluminium (Al) deposition. During a low temperature anneal at $500^\circ$C the Al was found to migrate throughout the entire device, even over micron length scales, forming an AlSi alloy device that conforms to the initial Si nanowire geometry. AlSi has a higher critical temperature (1.54~K) and critical magnetic field than Al allowing a broader range of  parameters to be studied. The resultant phase transformed nanowire structures are layered in geometry with a continuous Al-Si alloy wire sitting on the buried oxide of the SOI and a residual Si cap sitting on top of the wire. The phase transformed material is conformal with the predefined device patterns and the resultant structures are exceptionally smooth-walled. This behaviour was confirmed in five devices in two different batches. The main device here consists of a superconducting AlSi ring with a wire cross-section of $50 \times 50 \;\rm nm^2$ and an inner radius of 200~nm which produces periodic oscillations of the critical temperature $T_c$ as the strength of a magnetic field threading the ring is varied. This is consistent with the Little-Parks effect.\cite{Doll1961, Deaver1961, Little1962, Parks1964} The device has an extremely high quality factor as evidenced through the high critical current $I_C$ (2.8~$\rm A/cm^{2}$), large super-current/re-trapping-current hysteresis and current-voltage (IV) characteristics.

{\bf Device fabrication}

A schematic of the device fabrication process is shown in Fig.~\ref{Fig1}. We start with a SOI platform with $50$ nm Si on 150~nm SiO$_{2}$ (buried oxide, BOX) on bulk Si. Both the top Si layer and the substrate are Boron doped to a resistivity of $\rho=1-20$~$\Omega$.cm. The top Si layer was first degenerately doped using B ion implantation (6~keV, $1.7\times10^{15}\;\rm cm^{-2}$). The ion energy was chosen so that the B was confined to the top Si layer. The B concentration is expected to peak at 4$\times 10^{20}\rm\; cm^{-3}$. The implanted dopants were electrically activated with a 600$^{\circ}$C anneal for 10 minutes in a forming gas atmosphere. Recrystallization occurred via solid phase epitaxy. This dopant concentration is not high enough to induce superconductivity.\cite{Marcenat2010} 

After the dopant activation anneal, 50~nm wide nanowire devices consisting of rings were patterned with electron beam lithography (EBL). Inner radii of 100, 200 or 400~nm were patterned. The 200~nm rings are the focus of this work but preliminary results from the other devices with 100 and 400~nm inner radii rings are provided in Fig.~S2 of the Supporting Information. For the EBL, a 50~nm thick hydrogen silsesquioxane (HSQ) negative resist was used. 

Reactive ion etching (RIE) was employed to transfer the pattern into the top Si layer to form the nanowire device (Fig.~\ref{Fig1}(b)). An additional EBL step using 600~nm Poly(methyl methacrylate) (PMMA) resist allowed patterning of the contact pads. Immediately before electron beam evaporation of a $\sim$175~nm thick Al layer (Fig.~\ref{Fig1}(c)), a buffered hydrofluoric acid dip was employed for 4~seconds to strip the native oxide to ensure intimate contact between the Al and Si top layer. Alignment markers (5~nm titanium / 50~nm platinum) were used to align these two patterns. Finally, a thermal anneal at $450^\circ$C for 30~minutes under a forming gas atmosphere (Ar with a 5\% H content) was performed as depicted in Fig.~\ref{Fig1}(d). This step was discovered to induce Al migration throughout the entire device as discussed further below. \\

%%%%%%%%%%%%%%%
%%%%%%%%%%%%%%%
%%%%%%%%%%%%%%%

% \subsection{Structural characterisation}

\begin{figure*}
\begin{center}
\rotatebox{0}{\includegraphics[width=18cm]{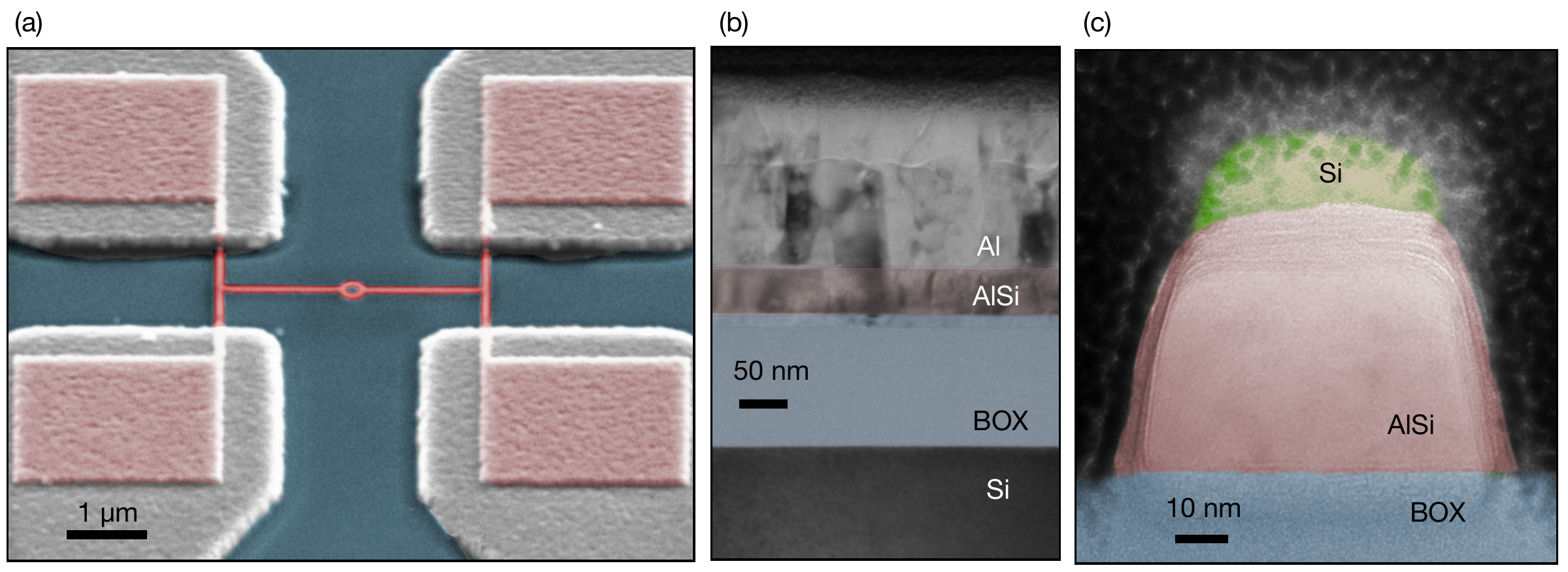}}
\end{center}
\caption[]{ (a) False-color SEM of the resultant device. The inner radius of the ring in this particular device is 100~nm. TEM cross-sections of (b) the contact pad region and (c) through the AlSi nanowire device. The colour coding, which indicates the different materials in the device, is informed by energy dispersive X-ray analysis measurements.} 
\label{Fig2}
\end{figure*}

{\bf Structural characterization}

% We first turn to the structural characterisation of the devices studied in this work. The resultant devices were assessed with secondary electron microscopy (SEM), transmission electron microscopy (TEM) and energy dispersive X-ray analysis (EDX). 
% The sample preparation for these XTEM measurements involved Pt deposition followed by focused ion beam milling of a lamella.

A false colour secondary electron microscopy (SEM) image of a fully fabricated $4$-terminal device with a ring of inner radius 100~nm is shown in Fig.~\ref{Fig2}(a). Fig.~\ref{Fig2}(b) and (c) show cross-sectional transmission electron microscopy (XTEM) images through the contact pad and through the nano-wire portion of a device, respectively. It is clearly observed that the Al has diffused throughout the device during the $450^\circ$C anneal, forming polycrystalline grains of AlSi alloy around the 10-50~nm size. The AlSi alloy conforms to the geometry defined by the original Si nano-wire and does not spread across the surface of the BOX layer exposed after RIE.

According to energy-dispersive X-ray analysis (EDX) the AlSi alloy contains approximately 13~at.\% Si. This is close to the Si content of the binary eutectic which is 11.7~at.\% silicon at the melting point of 577$^\circ$C.\cite{Murray1984} This is surprising since the anneal after Al deposition was 450$^\circ$C, well under the eutectic temperature. The long range transport of the Al may be influenced by the low resistivity Si and/or the constrained dimensions of the nanowire and requires further investigation. 

Another interesting feature of the AlSi wire clearly observed in Fig.~\ref{Fig2}(c) is the Si ribbon atop the AlSi wire. In this instance, the Si ribbon is approximately 10~nm thick. The presence of the Si on top of the AlSi suggests that the formation of the AlSi alloy may nucleate at the Si-BOX interface and this is a focus of ongoing studies on the details of this phase-transformation process. 

% Excess Si that exceeds the solid solubility limit of the alloy is presumably driven upwards by a AlSi alloy growth front. 

% The sidewalls are also coated with an oxygen rich material which is likely a native oxide. The presence of the native oxide in the fully alloyed devices suggests that full encapsulation with a device quality oxide should be possible. 

Multiple boundaries can be discerned around the perimeter of the wire in ~\ref{Fig2}(c) which is simply due to the TEM sampling a length of the nanowire. These faint boundaries have a greater variability on the top of the nanowire related to the tilt of the sample during measurement or a greater surface roughness on this facet. Importantly, this boundary appears to be very smooth. In contrast, Al is usually noticeably granular at these scales when deposited.\cite{Zhang:2000uc} The dark regions in these images represent the Pt deposited during TEM sample preparation.

AlSi alloys are known to display superior superconducting critical temperatures, $T_c$ to pure Al.\cite{Chevrier1987} Essentially, what is formed here is an Al-AlSi-Al superconducting nanowire where a ring is incorporated into the AlSi section. \\

% \subsection{Magnetoresistance characterisation}

{\bf Magnetoresistance characterization}

Electrical measurements as a function of temperature and perpendicular magnetic field, $B_\perp$ were performed with a cryogen-free, closed-cycle dilution fridge equipped with a superconducting vector magnet. The stability of the sample temperature was assessed to have a variance of $<0.3$~mK during measurement. 

The four terminal 200~nm radius ring device was wire bonded in a non-standard way. One bond broke so the remaining bond on the same side of the ring was double bonded. This means that pure Al is also addressed in the device in addition to the AlSi nanowire and ring. Results from devices with a standard bonding configuration are included in Fig.~S2 of the Supporting Information.

% \section{Results and discussion}

% \subsection{Magnetoresistance measurements}

\begin{figure*}
\begin{center}
\rotatebox{0}{\includegraphics[width=18cm]{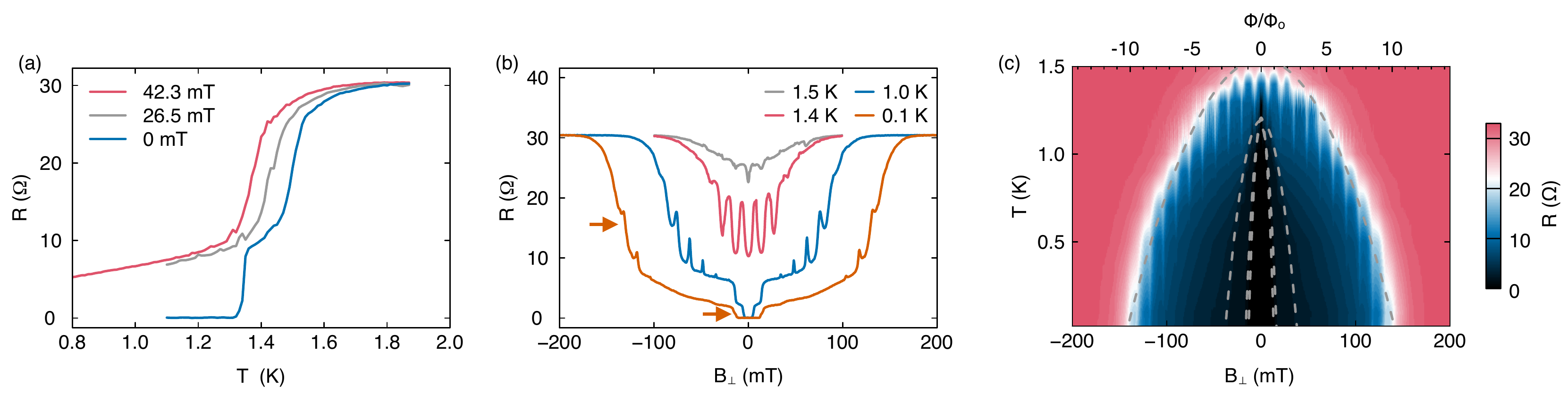}}
\end{center}
\caption[]{(a) The 200~nm AlSi ring $4$-terminal resistance as a function of temperature with different applied $B_\perp$ fields. (b) $B_\perp$ dependence of the resistance at various temperatures. Two arrows indicate the point of inflection of the main resistive steps on the 0.1~K data set. (c) The $B_\perp$ versus temperature with resistance color encoded. The dashed lines are fits to the resistance steps at half height using Eq.~\ref{tinkam}. The fluxoid period in units of  $\Phi/\Phi_{0}$, where $\Phi_{0}=13.87~\rm mT$ is shown on the top axis.} 
\label{Fig3}
\end{figure*}

Figure~\ref{Fig3} shows the 4-terminal lock-in resistance $R$ for the 200~nm radius AlSi ring device measured with a phase-synchronized lock-in amplifier (SR830). The modulation current for the lock-in was $1\;\rm \mu A$. The temperature dependence of $R$ is shown for three different $B_\perp$ values in Fig.~\ref{Fig3}(a). At $B_{\perp}=0$~T we observe two steps in resistance. The first at 1.54~K denotes the superconducting transition of AlSi. The second step at 1.34~K is where the transition occurs for the Al pads. For the two higher  $B_\perp$ values, only the AlSi-related transition is observed and the Al component of the device, in its normal state, provides a series resistance contribution.

The $B_\perp$ dependence at temperatures between 0.1-1.5~K is shown in Fig.~\ref{Fig3}(b). Prominent resistance oscillations are observable which are strongest close to the normal-superconducting phase boundary and have a period of $B_\perp=(n+1/2)\Phi_{0}$ where $\Phi_{0}=13.87\;\rm mT$.\cite{MTinkham75,Bezryadin2012}  From the relation $\Phi_\circ = h/2e = 13.87 \pi r^{2}$, this period gives an effective ring radius of $r=217.8$~nm which falls within the range of the nominal 200~nm and 250~nm inner and outer ring radii, respectively. For the other 100 and 400~nm inner radii rings we studied, we found values of 126 and 412~nm, respectively, again falling within the expected range (included in Fig.~S2 of the Supporting Information).

% This is where the total flux, which includes the external field as well as that formed from the persistent superconducting screening current (Meissner-Ochsenfeld effect) flowing through the ring's surface, is quantized in units of the fluxoid $\Phi_{0} = h/q$, where $q=2e$ is the fundamental charge pair within the superconductor.\cite{MTinkham75,Bezryadin2012} 

The shape of the oscillations changes with temperature and agrees with previous observations.\cite{Espy_2018} At the lowest temperature presented here, 0.1~K, the oscillations have a cusp-up shape. At 1.4~K the oscillations have a sinusoidal form. Finally, at higher temperatures they appear cusp-down in line-shape similar to what is observed for a SQUID.\cite{Espy_2018}

There are a number of resistance steps in the data presented in Fig.~\ref{Fig3}(b) indicated with the arrows for the 0.1~K data set. The step at 14.5~mT is associated with superconducting transition of the Al contact pad whereas the larger step around 135~mT corresponds to that of the AlSi component of the device. All resistance steps are well described by the equation:\cite{Tinkham1963,Moshchalkov1995}

\begin{equation}
T_c(B_\perp) = T_c(0)\left(1-\frac{\pi^2}{3}\left(\frac{w\xi B}{\Phi_{0}}\right)^2\right)
\label{tinkam}
\end{equation}

\noindent where the thickness is $w=50$~nm for these nano-wires. The magnetic field position of a step is defined at its half height and the fits using this equation are included in Fig.~\ref{Fig3}(c) (dashed lines). For the main resistance step that we associate with the AlSi rings, $\xi=161\pm2 \;\rm nm$ and $T_c = 1.54 \pm 0.03 \;\rm K$. The BCS superconducting gap for AlSi is then $234\pm5$~$\mu$V using $\Delta = 1.764 k_B T_C $.\cite{MTinkham75}

The inner step associated with the Al in the contact pads gives $T_c = 1.21\;\rm K$. The value of $w$ from this fit is not well-defined since the thickness of the contact pads contributing to this measurement is not known. Therefore, the associated $\xi$ value cannot be determined. 

Two other small resistance steps with fits shown in Fig.~\ref{Fig3}(c) could be discerned for this device between the Al and AlSi contributions with $T_c = 1.19$~K and 1.16~K and $\xi=0.60$ and $1.39\;\rm \mu m$, respectively. These steps are likely associated with the AlSi leads of the device and an Al-rich component close to the Al contact pads, respectively. However, other possible sources of resistance such as charge noise in the underlying oxide or non-equilibrium quasi-particles cannot be ruled out at this stage. Fig.~\ref{Fig3}(c) also contains an extrapolated data set of the resistance as a function of T and $B_\perp$. The oscillations are clearly observed here at half integer values of the fluxoid and the flux in units of $\Phi_{0}$ is denoted on the top axis. 

% $\Delta = 1.764 k_B T_C $ \cite{MTinkham75}

% The full data set measured over the range T = 0.5-1.5~K is provided in Fig.~S3 of the Supporting Information. 
%%%%%%%%%%%%%%%
%%%%%%%%%%%%%%%
%%%%%%%%%%%%%%%
\begin{figure}[h!]
\begin{center}
\rotatebox{0}{\includegraphics[width=7.1cm]{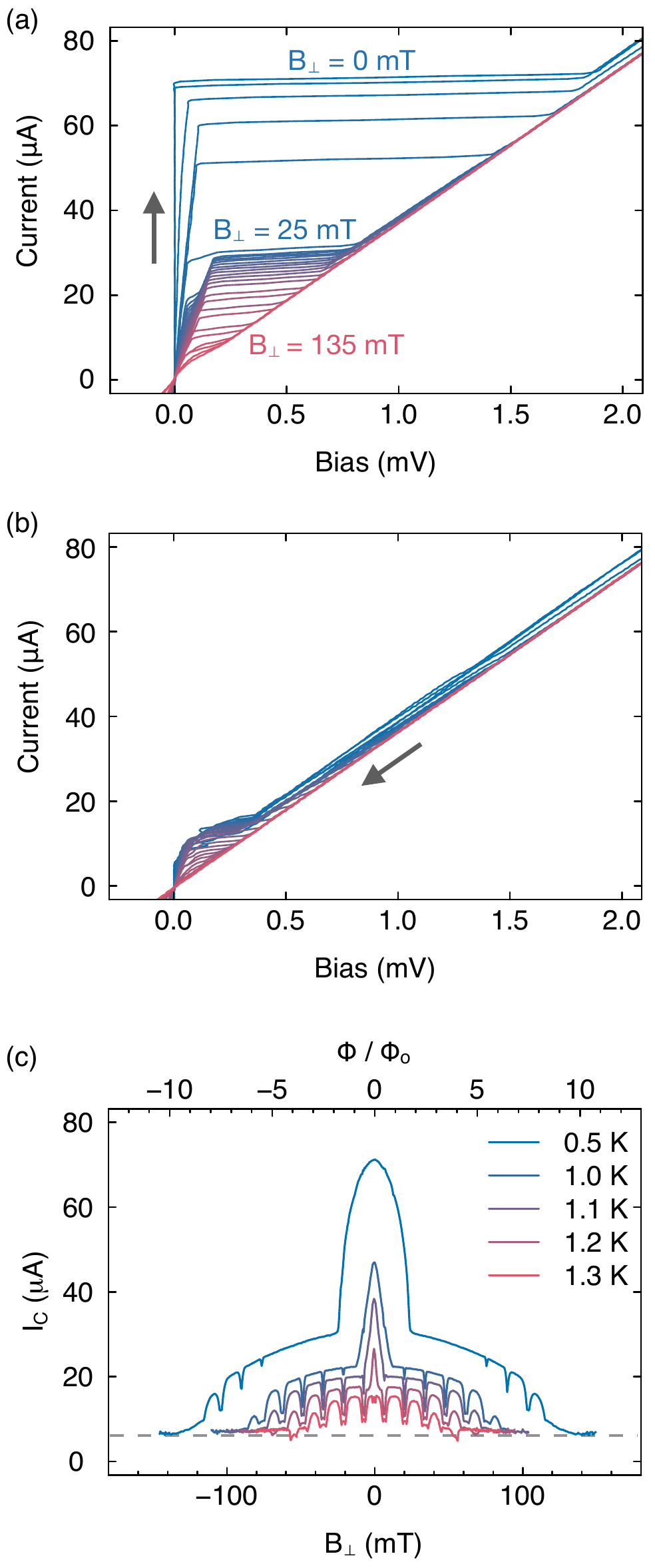}}
\end{center}
\caption[]{IV traces measured at 0.5~K for $B_\perp$ values between 0 and 135~mT in steps of 5~mT for (a) the super-current and (b) the re-trapping current. The current sweep direction is indicated with arrows. (c) Critical current as a function of $B_{\perp}$ extracted from IV data for temperatures between 0.5-1.3~K. No data is shown for $I<7\;\rm \mu A$ (dashed grey line) since reliable critical current values are more difficult to determine as explained in the text. The fluxoid values are indicated on the top axis.} 
\label{Fig4}
\end{figure}

Further insight can be gained via DC IV measurements. The IV traces were collected with a ramp generator to drive the current (SR DS345) between two contacts on opposite sides of the device, an amplifier to measure the bias voltage (SR560) between two additional contacts and an oscilloscope to digitize the output of the amplifier. A series of IV collected at 0.5~K is shown in Fig.~\ref{Fig4}(a) for $B_\perp$ values between 0 and 135~mT in steps of 5~mT. In these traces, only the increasing current ramps (supercurrent) are shown. The return currents, shown separately in Fig.~\ref{Fig4}(b), exhibit a re-trapping current at values much less than the critical current. This is due to self heating of the device at finite voltage values. The return current data will be presented in further detail below. 

% The ramp was a saw-tooth wave with a slope of ---~mV/s. 

In Fig.~\ref{Fig4}(a), the critical current at zero field is 70~$\mu$A ($2.8\times 10^{10} \;\rm A/m^2$). As the $B_\perp$ increases the Al pads go normal (from 10~mT) and contribute a resistive component ($\sim 2\; \Omega$) to the device resulting in an increasing slope in the IV trace close to $V=0$~V in Fig.~\ref{Fig4}(a). Another major resistive branch ( $\sim 6\; \Omega$) appears above $B_\perp = 25$~mT corresponding to the resistance steps indicated by the arrows in Fig.~\ref{Fig3}(b).

Figure~\ref{Fig4}(c) shows the critical current, $I_C$, as a function of $B_{\perp}$ for the 200~nm AlSi device at various temperatures. This was determined from the IV super-current traces like those displayed in Fig.~\ref{Fig4}(a). The trend as a function of $B_{\perp}$ displays clear oscillations with the same period as that observed with the lock-in method (Fig.~\ref{Fig3}). For $I<7\;\rm \mu A$ the $dV$ and $dI$ values are comparable to the noise so it becomes difficult to extract the switching current. As a consequence, the $I_C$ values do not reach below $I_C = 7\;\rm \mu A$ and the shape of some of the oscillations are not well defined, having a truncated dip.

%%%%%%%%%%%%%%%
%%%%%%%%%%%%%%%
%%%%%%%%%%%%%%%
\begin{figure*}
\begin{center}
\rotatebox{0}{\includegraphics[width=16cm]{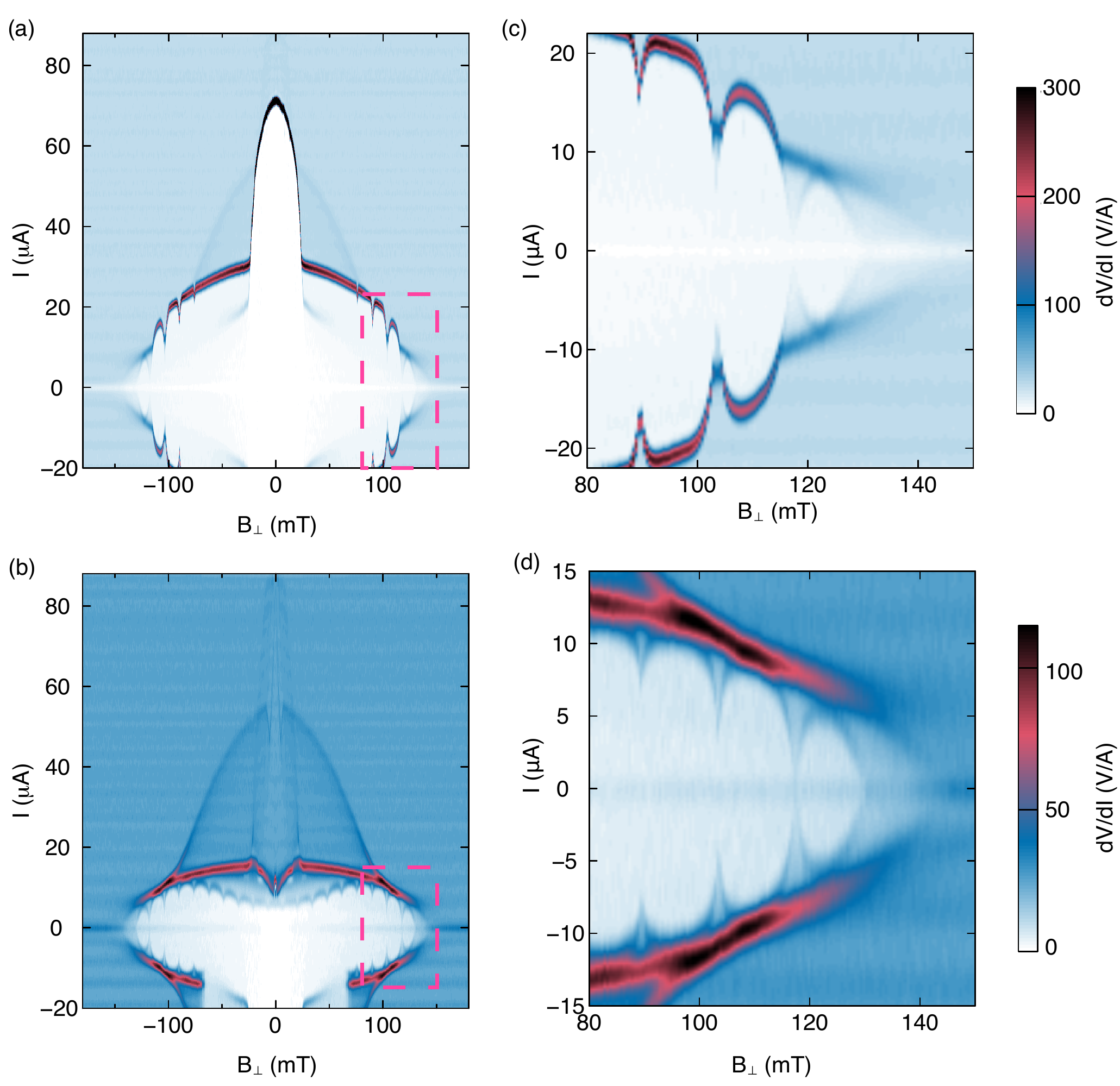}}
\end{center}
\caption[]{The 200~nm ring device $dV/dI$ versus $B_{\perp}$ determined from the IV measured at 0.5~K for (a) the super-current and (b) the re-trapping current. Close-ups indicated by the dashed boxes in (a) and (b) are shown in (c) and (d), respectively.}
\label{Fig5}
\end{figure*}

To further investigate the resistive branches appearing in the IV traces we plot $dV/dI$ in Fig.~\ref{Fig5}(a) and (b) for the super and return currents, respectively. In the normal phase $dV/dI=27\;\rm V/A$ for this device independent of field. Three dome-shaped components can be clearly discerned. The prominent central feature is associated with the superconducting phase of the Al contact pads. This feature denotes an operating region in which the device is fully superconducting, $dV/dI=0\;\rm V/A$ and is bound by large $dV/dI$ values where the current switches to the normal phase. The prominence of this feature is expected to be a result of the electrical bonding configuration employed. Indeed, this feature is less pronounced in data from other devices where a standard four contact van der Pauw configuration was used (Fig.~S2, Supporting Information). Furthermore, above 1.2~K, the critical temperature of Al, this feature disappears. The full data set consisting of $dV/dI$ plots for both the super and return currents over a 0.5-1.5~K temperature range are provided in Fig.~S3 of the Supporting Information for completeness.

Another broad dome-shaped feature extends to approximately $\pm 150 \;\rm mT$ and is convoluted with oscillations at half integer values of $\Phi_{0}$ from around 100~mT consistent with the resistance measurements of Fig.~\ref{Fig3}(b). This behaviour is also similar to a dc SQUID where the critical current is periodic with $B_\perp$ with the period inversely proportional to the area of the SQUID ring.\cite{MTinkham75} 

The two broad features shown in Fig.~\ref{Fig5}(a) arising from the Al and AlSi components of the device are similar to those reported recently for a Al-GeSi Josephson field effect transistor.\cite{Ridderbos:2020wo} This device was similarly formed by thermally inducing Al migration through a GeSi nanowire. 

The  $dV/dI$ plot in Fig.~\ref{Fig5}(a) also exhibits other more subtle features. For instance, there is a faint dome-shaped feature with a width of about $\pm 80 \;\rm mT$. These may arise from the accidental formation of Josephson junctions or AlSi material with a variable stoichometry. Both of these possibilities could  occur when AlSi is discontinuous and requires further study. 

% These are unique to this particular device. In a 400~nm ring device studied under similar conditions, for example, only the two dome-shaped boundaries that we attribute to Al and AlSi are observed (Fig.~S4 of the Supporting Information). ---- actually, there are faint features in the 400 nm map as well - but it cannot be seen unless the noise is low.

The return current $dV/dI$ also shows these features as displayed in Fig.~\ref{Fig5}(b). Interestingly, this includes the oscillations at half integer values of $\Phi_{0}$ in the return current. These are fainter than the super-current oscillations but appear over the full range of $B_\perp$ values within the AlSi superconducting phase. Furthermore, the return current oscillations are not superimposed over a domed shaped boundary as they are in Fig.~\ref{Fig5}(a). Instead, the envelope on which these oscillations are superimposed first increases in critical current to a maxima at $\pm$80~mT before decreasing again at higher fields. The oscillations are not usually observed in the return current since it is a phenomenon characteristic of rings in the superconducting phase.\cite{Bezryadin2012} However, there are some instances when oscillations are reported in the literature.\cite{Bruynseraede1996} In our case, we see that the device transitions to a superconducting state before the oscillations arise as indicated by changes in $dV/dI$ at current values greater than those where the oscillations appear in Fig.~\ref{Fig5}(b). This transition itself does not exhibit any oscillatory features. Therefore, although oscillations are observed in the return current, at least part of the device is already in a superconducting phase by this stage.

% The Al component of the device is not in direct contact with the ring. so Andreev reflections of normal electrons at the N/S boundary might not contain phase information from the ring. (\cite{Petrashov1993, Petrashov1995})

The return current behaviour may be a consequence of the Si ribbon that consistently forms on top of the AlSi nanowires during fabrication. This part of the device may not retain a stable superconducting state (via the proximity effect) and may therefore act as an efficient quasiparticle trap.\cite{Hosseinkhani2018}  As a consequence the trap would allow the normal current to freeze out more efficiently thereby improving the performance of the device. Further work is required to understand the impact of the Si ribbon on device operation. We also find that the geometry of the Si ribbon can vary between devices and possibly over a single device. This will result in a variation of the nanowire cross-section and thus variable superconducting parameters such as the critical current.  

% 

% the plots are slightly asymmetric about $I=0\;\rm A$ simply because the IV sweep range was asymmetric, between -20 and 90~$\mu A$, where transitions between $S$ and $N$ may not have occurred.

% The Si ribbon atop the device does not appear to act as a shunt resistor possibly because of its close contact with the AlSi material where it may also become superconducting via the superconducting proximity effect. The role of the Si ribbon will be discussed further below.

% Significantly, a feature of this device is that there is a temperature range (1.2-1.5~K) over which only the AlSi is in the $S$ phase and the Al is in the $N$ phase. Although Al-related features are observed here because of the wire bonding configuration employed, this observation nonetheless shows that devices and experiments can be designed to have parts of the device in the $N$ phase while maintaining $S$ in the AlSi portion. 

The shape of the oscillations is now considered with a focus on the regions displayed in Figs.~\ref{Fig5}(c) and (d). As discussed above, a transition from the normal state occurs before the oscillations appear in the return current. This behaviour can also be observed in the supercurrent oscillation at $B_{\perp}=104$~mT and 118~mT in (Figs.~\ref{Fig5}(c)). In this case the oscillations can also occur while part of the device remains in a superconducting phase with a resistive component. In contrast, at $B_{\perp}=90$~mT, the transition to the normal phase takes place at the apex of the oscillation (at $\pm 15.7\;\rm \mu A$).

%%%%%%%%%%%%%%%
%%%%%%%%%%%%%%%
%%%%%%%%%%%%%%%
\begin{figure}
\begin{center}
\rotatebox{0}{\includegraphics[width=8cm]{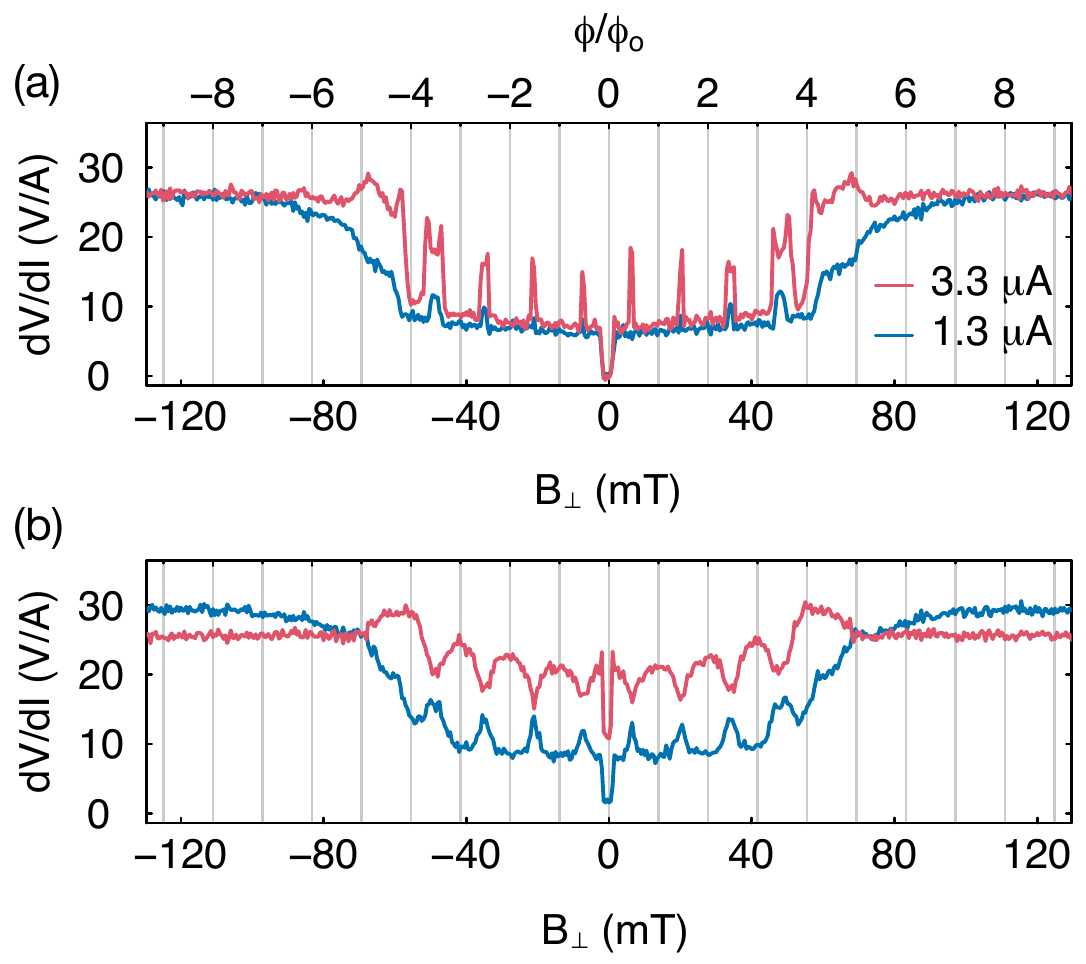}}
\end{center}
\caption[]{$dV/dI$ versus $B_{\perp}$ at 1.3~K for (a) the super-current and (b) the re-trapping current at measurement levels 3.3 and 1.3~$\mu$A. These were extracted from constant current line cuts through Fig.~\ref{Fig5} (a) and (b), respectively. The vertical lines denote the fluxoid values.} 
\label{Fig6}
\end{figure}

%%%%%%%%%%% 

As a consequence of the mixed superconducting and resistive states, the shape of the oscillations can be highly variable. This will have an impact on how the device might be used for magnetometry applications for example. To illustrate this variability in shape, $dV/dI$ is plotted for two fixed bias currents applied while the device was at 1.3~K in Fig.~\ref{Fig6}. At this temperature the Al is in the normal state. Both the super- and the return currents $dV/dI$ are displayed in Fig.~\ref{Fig6}(a) and (b), respectively. For the super-current $dV/dI$ (Fig.~\ref{Fig6}(a)) very sharp peaks appear at $(n+1/2)\Phi_\circ$. The width of these peaks is around 1~mT but there is an apparent broadening as $B_{\perp}$ increases. Furthermore, the peak at $\pm 48 $~mT is observed to split simply due to a resistive state transition of the device. 

For the re-trapping dV/dI (Fig.~\ref{Fig6}(b)) the current can be chosen to either display standard ($I=1.3\;\rm \mu A$) or inverted ($I=3.3\;\rm \mu A$) oscillations, both with the $(n+1/2)\Phi_\circ$ period. With the appearance of a number of resistance steps the device can exhibit a broad range of complex oscillatory features. This is likely due to the presence of compositional inhomogenities along
the nanowire which may lead to variations in superconducting properties.

% These could indeed appear to have an effective $h/4e$ period. 
% \cite{Petrashov1993,Zadorozhny2001} 

% \section{Conclusion}
In conclusion, a superconducting device was fabricated on a SOI platform using standard semiconductor industry processes. Al from electrical contact pads was found to migrate throughout a pre-patterned Si nanowire device, even over micron length scales, forming an AlSi alloy. The phase transformed material is conformal with any predefined device patterns and the resultant structures are exceptionally smooth-walled. Excess Si resides on the top surface of the resultant AlSi nanowire structure suggesting that the Al-Si interactions preferentially take place at the buried oxide - Si interface. This resulted in a unique Si ribbon structure which covered the top part of the device. 

The temperature and magnetic field dependence of AlSi nano-ring devices displayed periodic features in the differential resistance as well as the critical current. These oscillations are a result of fluxoid quantization. The re-trapping current also exhibited oscillations which may arise from efficient quasiparticle trapping in the Si ribbon. 

Finally, with greater control over the Al migration process it may be possible to form junctions between AlSi and Si and to explore the possibility of Josephson tunnelling in a configuration that is different to the conventional thin insulating barrier Josephson junctions.

% This allows the device to reach a superconducting phase efficiently in a regime where the oscillations can be observed. In this case, the oscillations occur while the device is in a mixed superconducting-resistive phase. Similar behaviour is also observed in the super-current under high applied magnetic fields. 

% Finally, we expect new, high precision experiments will be enabled by this improvement in superconducting device fabrication. Furthermore, the ability to fabricate superconducting components on Si offers the possibility to form junctions between the two materials and to explore the possibility of Josephson tunnelling in a configuration that is different to the conventional thin insulating barrier Josephson junctions. 

\section{Acknowledgements}
This work is funded by an Australian Research Council (ARC) grant (DP200103233). We acknowledge the ARC Centre of Excellence for Quantum Computation and Communication Technology (CE170100012) for financial support. The AFAiiR node of the NCRIS Heavy Ion Capability is also acknowledged for access to ion-implantation/ion-beam analysis facilities. M.S. acknowledges the Melbourne international fee remission scholarship (MISRF) and the Melbourne international research scholarship (MIRS). D.L.C. is supported by ARC grant DP190102852. J.H.C. is supported by the ARC Centre of Excellence FLEET (CE170100039) and the Australian National Computational Infrastructure facility. Finally, we thank Stephen Gregory for technical support.

% \bibliography{bib}

%merlin.mbs aipnum4-1.bst 2010-07-25 4.21a (PWD, AO, DPC) hacked
%Control: key (0)
%Control: author (8) initials jnrlst
%Control: editor formatted (1) identically to author
%Control: production of article title (0) allowed
%Control: page (1) range
%Control: year (1) truncated
%Control: production of eprint (0) enabled
%

\newpage
\clearpage
% \appendix
\renewcommand{\thesection}{S\arabic{section}}
\renewcommand\thefigure{S\arabic{figure}}    
\renewcommand{\baselinestretch}{1.5}

\pagenumbering{roman}
\setcounter{page}{1}
\setcounter{section}{0}
\setcounter{figure}{0}

\onecolumngrid{

{\bf\large \noindent Supporting Information:  \\Phase transformation-induced superconducting aluminium-silicon alloy rings}
\vspace{3mm}

{\noindent B. C. Johnson,$^1$ M. Stuiber,$^2,3$ D. L. Creedon,$^2$ A. Berhane,$^4$ L. H. Willems van Beveren,$^2$ S. Rubanov,$^5$ J. H. Cole,$^6$ V. Mourik,$^4$ A. R. Hamilton,$^4$ T. L. Duty,$^4$ and J. C. McCallum$^2$}\\
\noindent {\em $^{1)}$Centre of Excellence for Quantum Computation and Communication Technology, School of Engineering, RMIT University, VIC, 3001, Australia\\
$^{2)}$School of Physics, University of Melbourne, VIC 3010, Australia\\
$^{3)}$Melbourne Centre for Nanofabrication, VIC 3168, Australia\\
$^{4)}$School of Physics, University of New South Wales, Sydney, NSW, Australia\\
$^{5)}$Ian Holmes Imaging Centre, Bio21 Institute, University of Melbourne, VIC, 3010, Australia\\
$^{6)}$School of Science, RMIT University, VIC, 3001, Australia
}

\section{Structural characterisation}
Figure~\ref{S1}(a) shows a cross-sectional transmission electron microscopy (TEM) image of an AlSi nanowire fabricated by inducing Al migration through a silicon nanowire formed on a silicon on insulator platform. Energy dispersive X-ray analysis (EDX) is used to identify the composition of each component of the device. Fig.~\ref{S1}(b) shows the oxygen content is slightly higher in the top ribbon layer of the wire. Fig.~\ref{S1}(c) shows the silicon and aluminium components. The silicon content in the core of the wire is about 13~at.\% (i.e. Al$\rm _{1-x}$Si$\rm _x$, x=0.13). We suspect that the AlSi alloy formation nucleates at the Si/SIO$_2$ interface and proceeds epitaxially while consuming Si. Excess Si, above the solubility limit in AlSi is plowed ahead of the AlSi-Si interface during growth. 

The growth mechanism is distinct from the case of free standing SiGe nanowires where the Al reaction proceeds along the length of the wire.\cite{Luong2020} This is possibly because of the presence of the Si/SIO$_2$ interface.

% The Si rich top ribbon appears to be slightly thicker at 15~nm for this particular wire, whereas the wire in Fig.~2 of the main text had a thickness of 10~nm. It is not clear at present whether this is due to the tilt of the sample during measurement or to an actual thickness variation. The latter could arise from variations in the equilibrium concentration of Si in Al at the anneal temperature. 

\begin{figure*}[h]
\begin{center}
\rotatebox{0}{\includegraphics[width=17cm]{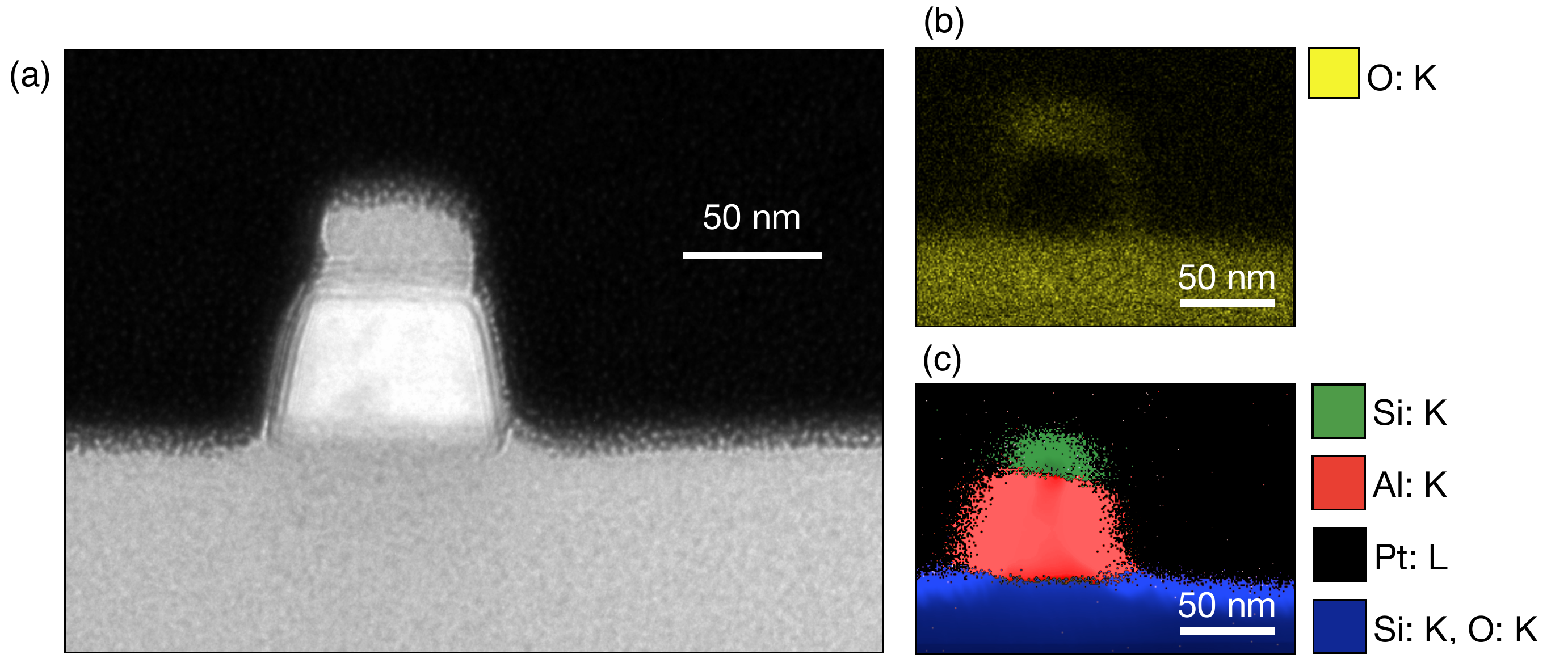}}
\end{center}
\caption[]{(a) Cross-sectional transmission electron microscopy (TEM) image through a single AlSi alloy nanowire. (b) The corresponding energy dispersive X-ray analysis image of the oxygen and (c) the silicon, aluminium and oxygen components within the nanowire.} 
\label{S1}
\end{figure*}
 
%%%%%%%%%%%%%%%
%%%%%%%%%%%%%%%
%%%%%%%%%%%%%%%
\section{Magnetoresistance measurements}

Figure~\ref{S2} shows the critical current as a function of applied magnetic field for devices with 100, 200 or 400~nm inner radii rings. The 100 and 400~nm devices were wire bonded in a standard van der Pauw configuration. One of the bonds on the 200~nm device broke during initial measurements. To progress with the measurement we double bonded one of the contacts. As a consequence, the central feature which is due to thick Al in the contact pads is much more prominent in this device than the 100 and 400~nm devices.

Each device exhibits critical current dips at $B_\perp=(n+1/2)\Phi_{0}$ as the perpendicular magnetic field, $B_\perp$ is varied with $\Phi_{0}=41.41$, 13.87, and 3.88~mT for the 100, 200 and 400~nm rings, respectively. From the relation $h/2e = \Phi_{0} \pi r^{2}$,\cite{MTinkham75,Bezryadin2012} these periods give an effective ring radii of $r=126.1$, 217.8 and 411.7~nm which falls within the range of the nominal inner and outer ring radii. Data for the 100~nm ring in Fig.~\ref{S2}(a) includes measurements with the addition of a parallel field, $B_{||}$ in the range 0-140~mT. As $B_{||}$ increases the critical current dips become deeper.

% 411.7173, 217.8435, 126.0754 # nm
% 3.883e-3, 13.87e-3, 41.41e-3 # T

\begin{figure*}
\begin{center}
\rotatebox{0}{\includegraphics[width=17cm]{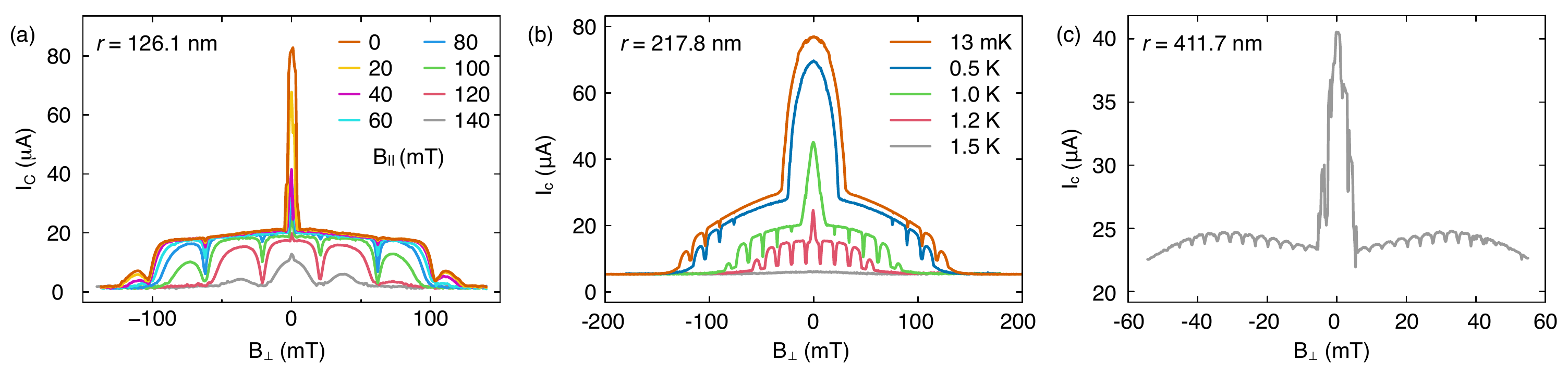}}
\end{center}
\caption[]{Critical current for AlSi ring devices with nominal inner ring radii of (a) 100, (b) 200 and (c) 400~nm. 
} 
\label{S2}
\end{figure*}

\section{Temperature dependence}

The $dV/dI$ plots extracted from the current-voltage traces for the super and return currents of the 200~nm device are displayed in Figs.~\ref{S3} (a) and (b), respectively. The data were collected over a range of $B_\perp$ and temperatures. The critical current of the central dome-shaped feature between $\pm 10\;\rm mT$  decreases significantly as the temperature increases. It completely disappears between 1.2 and 1.3~K consistent with the $T_c$ of pure Al. Likewise, the faint dome-shaped features which are clearly observed in Fig.~\ref{S3}(b) also disappear by a similar temperature. 

Clear oscillations with $B_{\perp}$ can be discerned for both the super and return currents at each temperature studied.

\begin{figure*}
\begin{center}
\rotatebox{0}{\includegraphics[width=17cm]{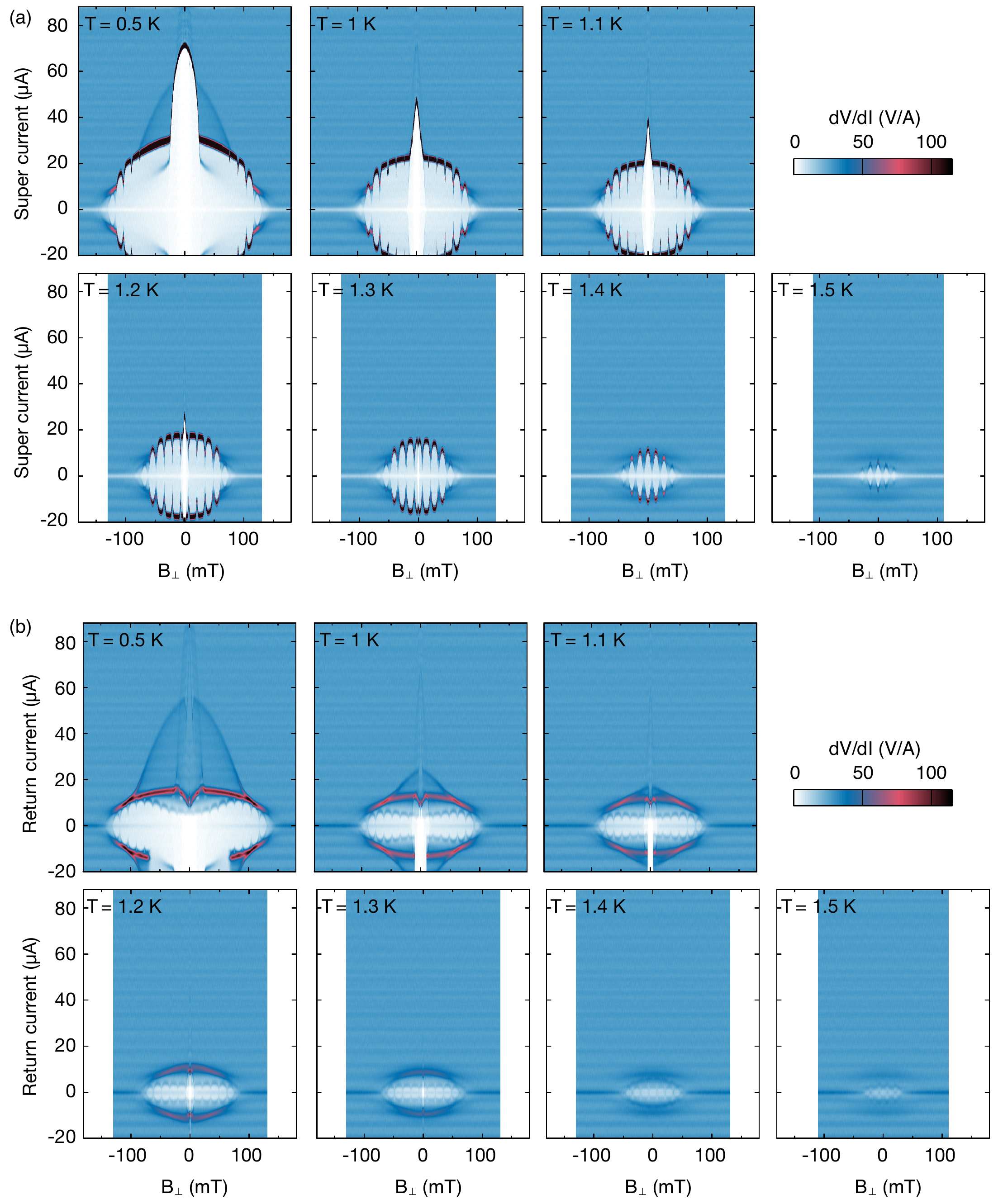}}
\end{center}
\caption[]{$dV/dI$ versus $B_{\perp}$ for the 200~nm radii ring  for (a) the super-current and (b) the re-trapping current at temperatures of 0.5-1.5~K. These were determined from current voltage measurements.} 
\label{S3}
\end{figure*}

 }

\end{document}